\documentclass[12pt,preprint]{aastex}
\received{}
\accepted{}
\slugcomment{}
\newcommand{\etal}{et al.}
\newcommand{\fe}{Fe~K$\alpha$}
\newcommand{\kev}{keV}
\newcommand{\mcg}{MCG-6-30-15}
\begin{document}

\title{The sharp drop in the flux striking the Accretion Disk}

\author{Huang Wei\altaffilmark{1},
}

\begin{abstract}
In this paper,We present a simple relativistic approach to analyze
the flux striking the disk which is possibly from a source up the
Black hole.The X-ray source is locate above an accretion disc
orbiting around the black hole,this assumption is invoked by
recent studies about iron k$\alpha$. we compute and arguing that
due to the light bending effect near black hole ,the flux striking
the disk surface may be very concentrated,which will undoubtable
change the disk's ionization state hence change the iron line's
ionization state and emissivity.Also,Our model may explain the
steep power law when modelling the lines.
\end{abstract}

\altaffiltext{1}{Center for astrophysics, University of Science
and Technology of China,Hefei,Anhui,230026 P.R.China;
huangwei@mail.ustc.edu.cn}

\section{Introduction}

\label{sect:intro} With the discovery of \fe\ emission at
6.4~\kev\ and spectral hardening at 20--30~\kev\ in the X-ray
spectra of Seyfert~1 galaxies (Pounds \etal\ 1990), it was a
center figure in probing the physics around black holes.ASCA's
observation about the \mcg 's board and skewed profile reveal two
important physical mechanisms-the gravitational and the Doppler
effect.There are also other sources which show board and skewed
profile.Various explanation and observation about the \fe have
been presented( see fabian 2000,Reynolds \& Nowak 2003 for
details.)Another intriguing thing is that,the sources always show
a X-ray continuum with power-law distribution ,till now,there is
no convincing explanation about the production of the X-ray
continuum.

    From both the observational and theoretical  side,the iron line is expected to be
    the reflection from the accretion disk .A lot of computational method to compute the iron line profile
from the vicinity of black hole has been invited(fanton
1997,Bromley 1994,labor 1991),among which the most renowned is the
ray-tracing method due to it's efficiency in computing.The
differences in the line profiles from source to source have
possibly implied variation in the geometry,the view angle, the
structure of the BH-accretion disk system.With XMM's high S/N data
,it greatly facilitate the astronomical community to study the
iron k$\alpha$ in AGN. In order to model the \fe many detailed
calculations about the accretion disk must be performed.Many
author have make computation considering the geometry ,such as
flat ,concave,conical,thick.(hartnoll 2002),Merloni founded that
the inner boundary condition of accretion flows have a significant
effect on the profile. The ionization state and fe abundance in
the accretion disk also have been studied by Ballantyne
1999,fabian 2002.Nayakshin's two-layer model in the surface of
accretion disk also seems plausible.Recent MHD studies have
proposed that there may be exist spiral velocities in accretion
disk which will brought quasi-humps in the profiles.

    However,for simplicity,the disk emissivity is usually assumed
    to have a power-law profile $F(r)\simeq r^{-\alpha}$(where $r
    =R/R_s=R~c^2/2GM$ is the distance from the center in units
    of schwarzshild radii),and the index of such power-law is
    generally assumed to be given by the accretion disc emissitvity
    index in a standard geometrically thin and optically think
    disk(Shakura Sunyaev 1973).The fact that both wilms \etal 2001
    and Miller \etal 2002 require a larger value of $\alpha$ to fit
    the line profile has prompted speculation that,in the inner
    disc disc region,additional energy dissipation must be taking
    place.The long MCG-6-30-15's observation with XMM-newton reported by
    fabian \etal 2002 show that,the strong ,skewed iron line is
    clearly detected and is well characterized by a steep emissity
    profile within 6GM/c$^2$ and a flatter profile beyond.

    In contrast to explain the steep profile in the non-zero
    torque theory,which needs a comprehensive study of the MHD
    in accretion disk.we just make the simplest computation to deduced
    the flux striking the surface of the disk in the GR
    scheme,hence see the effects on line profiles,and the result
    is very useful in accounting for the steep emissivity in the
    inner part of the disk.
\section{The model and computation}

\label{sect:cdens} Numerical models of reflection spectra from
accretion discs began with the simplest case: assuming a static,
neutral and constant density slab of material irradiated by a
power-law continuum of X-rays. George \& Fabian (1991) and Matt,
Perola \& Piro (1991) performed Monte-Carlo calculations of the
reflection spectra in such circumstances, and provided predictions
on the equivalent width (EW) of \fe\ for different reflection
geometries.In this paper,we also assume the source is centredly
located,static with a power law flux $I_{\nu}=A~\nu^{-\alpha}$,and
this assumption is very close to the observation about the X-ray
continuum.

the gravitational field of a rotating black hole is described by
the Kerr metric(we use units c=G=1)
\begin{equation}
d~s^2=-(1-\frac{2Mr}{\Sigma})d~t^2-\frac{4M~a~r}{\Sigma}\sin^2\theta~dtd\phi
+\frac{A}{\Sigma}\sin^2\theta~d\phi^2+\frac{\Sigma}{\Delta}dr^2+\Sigma~d\theta^2
\end{equation}
where
\begin{equation}
\Sigma=r^2+a^2\cos^2\theta\\
\end{equation}
\begin{equation}
\Delta=r^2+a^2-2M~r\\
\end{equation}
\begin{equation}
A=(r^2+a^2)^2-a^2\Delta\sin^2\theta\\
\end{equation}

We can see that the metric depends only on two parameters,the mass
$M$ of the black hole and the specific angular momentum $a=J/M$.
And it is straight forward to formulate that the photo in Kerr
metric should has the relation in different place.We assume the
photo propagate from place$(r_1,\theta_1)$to place$(r_2,\theta_2)$
then we can get
\begin{equation}
\frac{\nu_1}{\nu_2}=(\frac{g_{00}(r_1,\theta_1)}{g_{00}(r_2,\theta_2)})^{1/2}
\end{equation}
and due to the invariance in general relativity
\begin{equation}
\frac{I_{\nu}}{\nu^3}=const
\end{equation}
so we can deduce that
\begin{equation}
\frac{A_1}{\nu_1^{3+\alpha}}=\frac{A_2}{\nu_2^{3+\alpha}}\label{changeA}
\end{equation}
notice here,in the photons prorating in the Kerr metric,the A is
changing,due to the frequency changing.so we can get
\begin{equation}
A_2=A_{20}~(\frac{g_{00}(r_1,\theta_1)}{g_{00}(r_2,\theta_2)})^{(3+\alpha)/2}\label{intense}
\end{equation}
as we know,in the newtonian space-time the intensity of the flux
should not be changed,how ever,Eq. ~\ref{intense} show that the
intensity changed due to a factor.The same,we can get
\begin{equation}
F_{bending}=F_{straight}~(\frac{g_{00}(r_1,\theta_1)}{g_{00}(r_2,\theta_2)})^{(3+alph)/2}
\label{bending}
\end{equation}
where $F_{bending}$is the flux striking the disc in Kerr
metric,while$F_{straight}$is the fomat in newtonian space-time. In
the newtonian space-time.In the model of point like,static source
up the black hole,the flux striking the disk should be
\begin{equation}
F_{straight}=\frac{L~h}{4\pi~(h^2+r^2)^{3/2}}
\end{equation}
in Kerr black hole,we know that
\begin{equation}
g_{00}=-(1-\frac{2M~r}{r^2+a^2~\cos^2\theta})
\end{equation}
we set r,a,in unit of m,then$F_{straight}$can have the detailed
formula
\begin{equation}
F_{bending}=\frac{L~h}{4\pi~(h^2+r^2)^{3/2}}~(\frac{1-\frac{2h}{h^2+a^2}}{1-2/r})^{(3+\alpha)/2}
\end{equation}
for a non-rotating black hole,we can get
\begin{equation}
F_{bending}=\frac{L~h}{4\pi~(h^2+r^2)^{3/2}}~(\frac{1-2/h}{1-2/r})^{(3+\alpha)/2}
\end{equation}
when we set r,h in unit of Schwichild radii $2M/r$,the formula
change to
\begin{equation}
F_{bending}=\frac{L~h}{4\pi~(h^2+r^2)^{3/2}}~(\frac{1-\frac{h}{h^2+a^2}}{1-/r})^{(3+\alpha)/2}\label{qin}
\end{equation}
from Eq.~\ref{qin} we can see that the flux striking the accretion
disc has a close relation with both the photo index $\alpha$, the
accretion rate $a$,and the sources location.Observations shows
that the source X-ray continuum always have a photo index
$\alpha\sim 2$.

when we compute the $F_{bending}$ in this model,we find that in a
extremely rotating kerr metric the flux concentrated on only a
very small region near the center ,while other place left nearly
no illumination at all.It is apparently a GR effect.when it comes
to the schwchild metric,the power law approximation is also not
good to describe the emissivity,but it is rather flatter to the
case in Kerr Metric.We can conclude that the need of steep power
law model may due to a rapidly rotating black hole.

\section{Discussion}

our work is invoked by recent report of steep emissity in
    inner part of accretion disk(fabian 2002).In the observation
    of MCG-6-30-15,Fabian use a broken emissivity law to model
    the data,and with good result.However,they assume that the
    steep emissivity is a result from two reflector in the disk.
    In our work we can see that,the steep emissivity may come from
    the effect of light bending near the center black hole.And the
    concentrating of flux,will give inner side a rather high
    ionization parameter,$\xi$,
    which may result a line component at
    $\sim6.8$keV.There are also other sources,such as IRAS
    13224-3809(Boller 2003),which should induce a steeper
    emissivity law.The steep emissivity law may be a evidence of
    a centering rapidly rotating black hole.

    In present X-ray study,there are lots of uncertainty in modelling
the iron lines in AGN,the S/N is a big problem to determine the
real shape of the line.Former observation confirms that there
exist many AGN which contain board and skew iron line,indicating
that the line is produced very near the black hole.however,recent
XMM observation show that there does not ubiquitous exist board
iron line,instead,a narrow cole at 6.4 keV is present.This has not
been included in this paper.

    our work show that in the GR effect,the emissity can not
    be simply regarded as a power law as a function of r,instead,
    it should be described by various factors such as a,the source's position
    the photo index.When computed in the GR
    we find in the kerr metric the large concentration of the flux in the accretion
    disc,rather left other part of the disk "empty",also,the large
    concentration in flux will undoubtedly invoke the model of ionized disk
    describe by Ross 1999.And our work is only a simplified model
    to test the GR effects,further work such as the disk motion
    and density under high illumination which may affect the
    emissity should be invoked.

\acknowledgements

HW thanks for useful discussion with colleges in AGN research lab
in CFA

\clearpage
\begin{figure}
\centering
\includegraphics[width=150.0pt,height=150.0pt,angle=270]{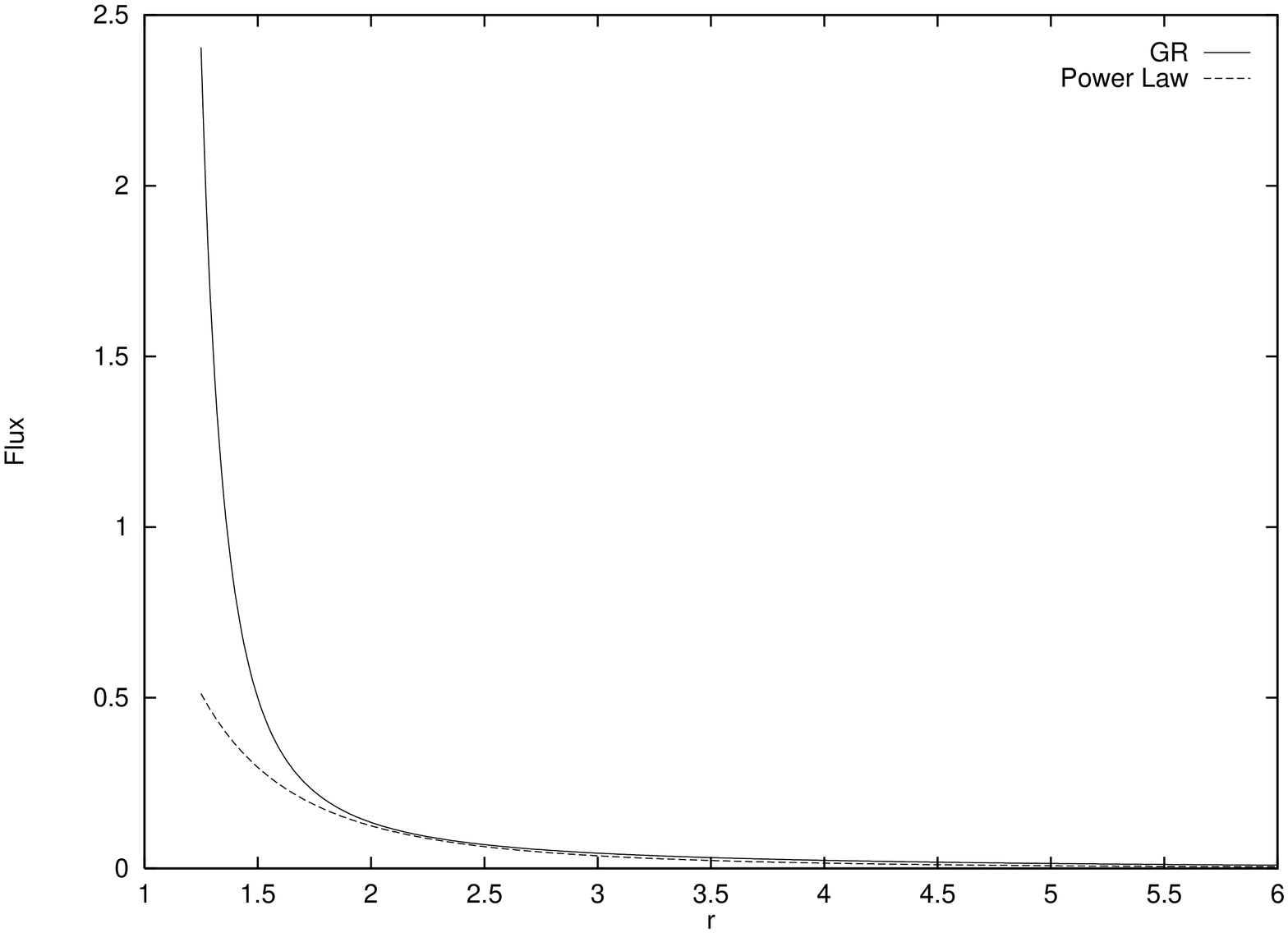}
\caption{the flux striking the disk as a function of r,and a.In
this picture,we set h=5
a=0.998,$\alpha$=3,$r_{in}=1.24,r_{out}=6$} \label{1}
\end{figure}
\begin{figure}
\centering
\includegraphics[width=150.0pt,height=150.0pt,angle=270]{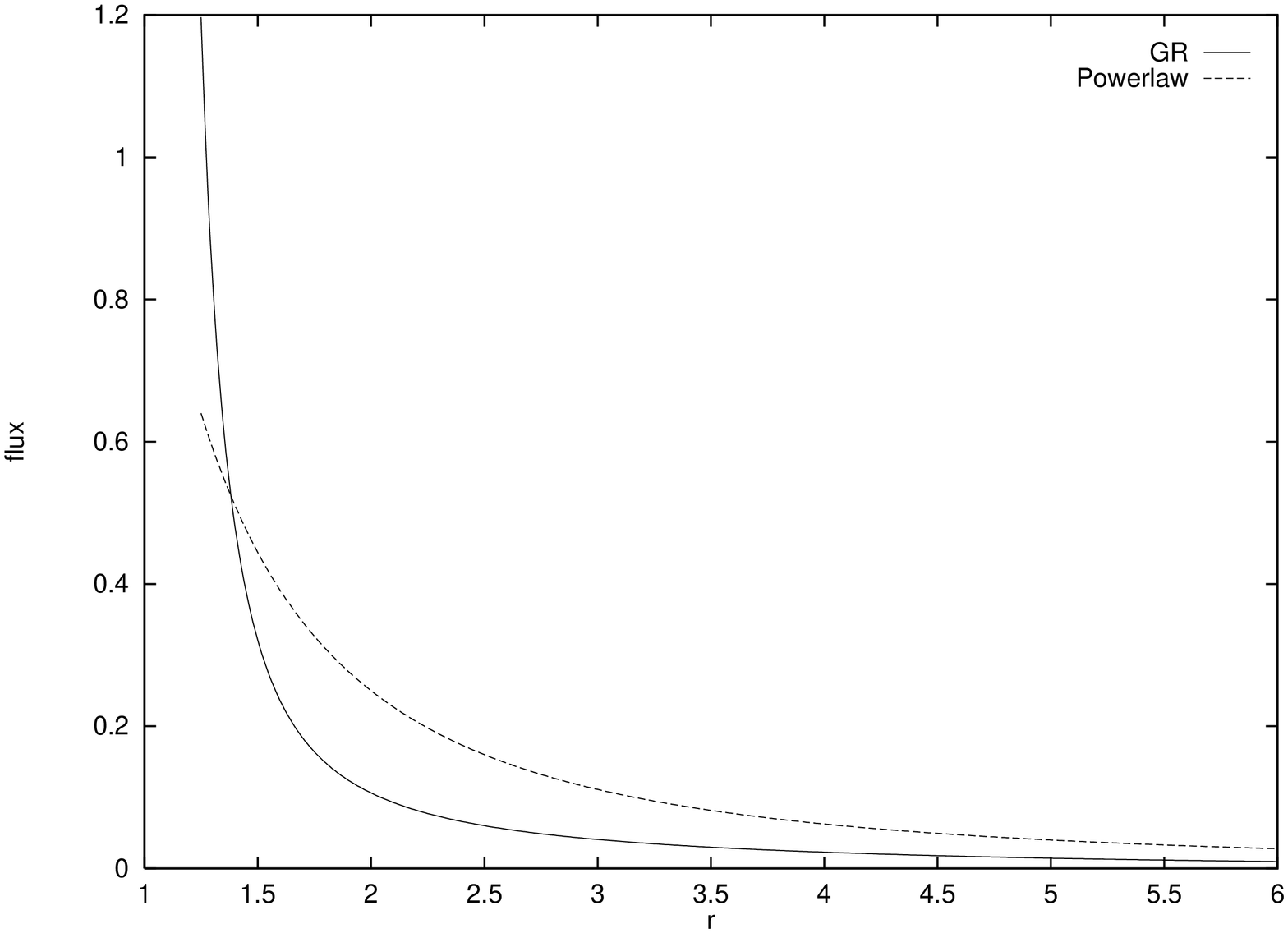}
\caption{the flux striking the disk as a function of r,and a.In
this picture,we set h=5
a=0.998,$\alpha$=2,$r_{in}=1.24,r_{out}=6$} \label{2}
\end{figure}
\begin{figure}
\centering
\includegraphics[width=150.0pt,height=150.0pt,angle=270]{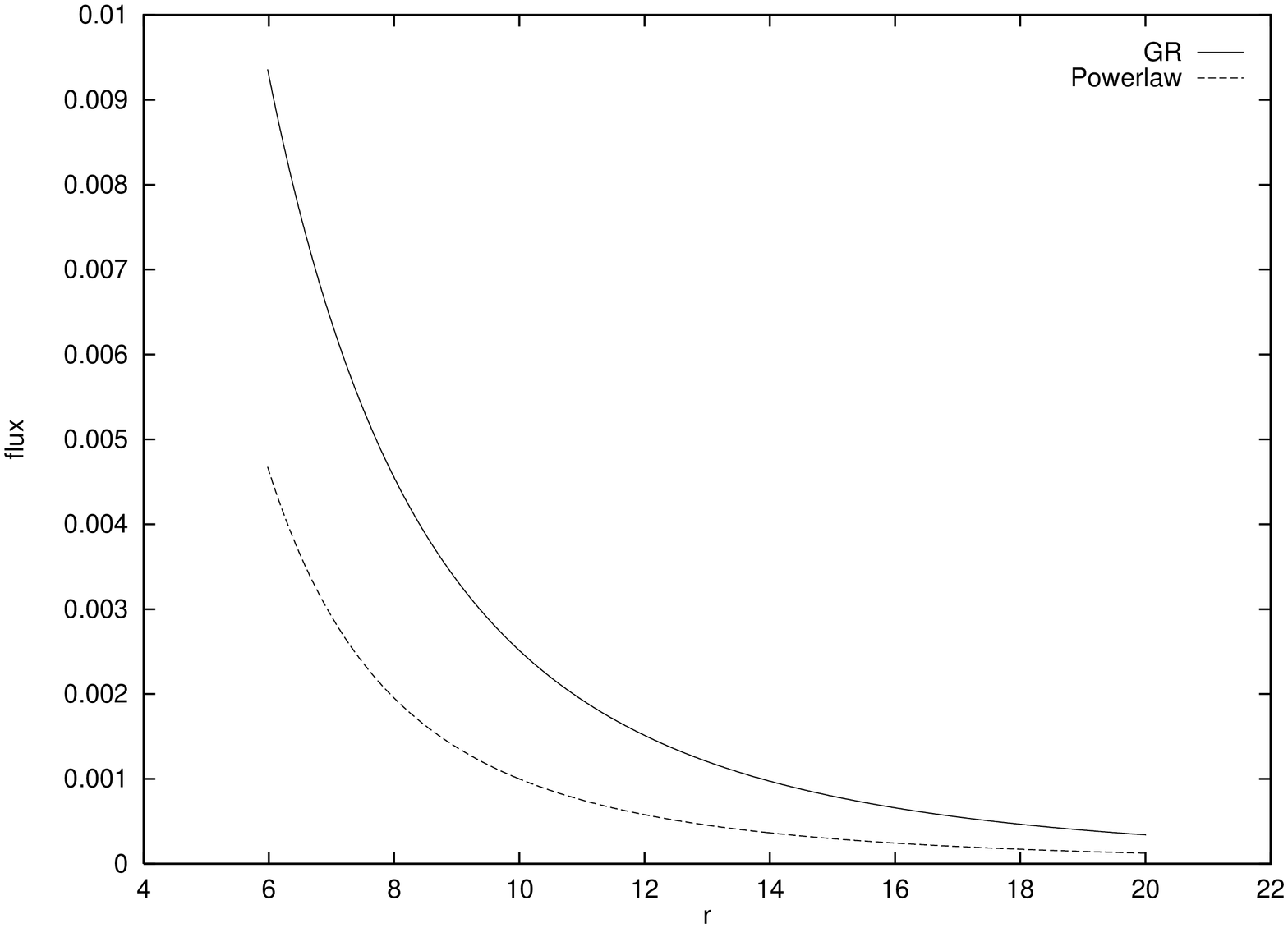}
\caption{the flux striking the disk as a function of r,and a.In
this picture,we set h=5
a=0.01,$\alpha$=3,$r_{in}=5.97,r_{out}=20$} \label{3}
\end{figure}
\begin{figure}
\centering
\includegraphics[width=150.0pt,height=150.0pt,angle=270]{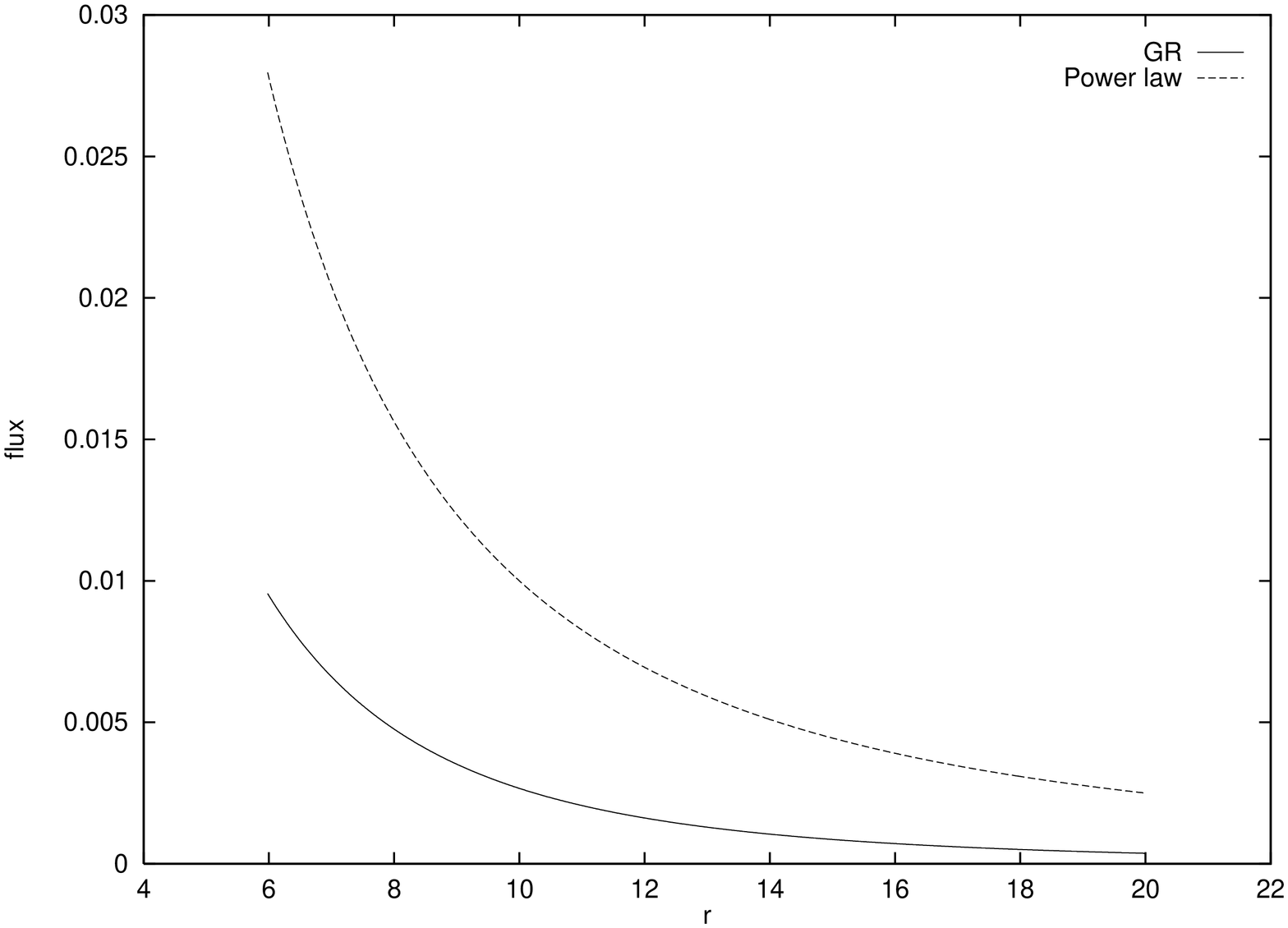}
\caption{the flux striking the disk as a function of r,and a.In
this picture,we set h=5
a=0.01,$\alpha$=2,$r_{in}=5.97,r_{out}=20$} \label{4}
\end{figure}

\end{document}